\documentstyle[12pt,kolnik,epsfig]{article}
\begin{document}
\begin{titlepage}
\begin{flushright}
PRA-HEP 95/9\\
September 1995\\
hep-ph 9603321\\
\end{flushright}
\vskip1,5cm
\begin{center}
{\huge \bf Electroweak interactions\\
 \vskip0,5cm
  and high-energy limit}
\footnote {Lecture notes for Triangle Graduate School 95
in Particle Physics, 2nd Austrian School held at
Bundesheim Raach, September 25-30, 1995}\\
\vskip1cm
{\bf An introduction to Equivalence Theorem}
\vskip3cm
J. Ho\v{r}ej\v{s}\'{\i}\footnote{e-mail: {\tt jiri.horejsi@mff.cuni.cz}}\\
\vskip0,5cm
{\it Nuclear Centre, Faculty of Mathematics and Physics\\
Charles University, V Hole¨ovi‡k ch 2, 180 00 Prague 8,
Czech Republic}

\vspace{2cm}

\noindent
{\bf Abstract}
\end{center}

\noindent
A pedagogical introduction to the equivalence theorem for
longitudinal vector bosons in electroweak theories is presented
and the problem of high-energy behaviour of scattering amplitudes
in the Standard Model is briefly reviewed. To make the treatment
self-contained, the basics of the Standard Model are summarized
in an appendix.

\end{titlepage}
\newpage

\begin{titlepage}
\tableofcontents
\end{titlepage}

\newpage

\section{Introduction}

Long since it has been known that in theories involving charged vector bosons
with non-zero mass (i.e. typically in weak interaction models) the
tree-level Feynman diagrams may diverge badly in the high-energy limit.
The divergences are associated with physical states of spin -1
particles carrying longitudinal polarization, i.e. zero helicity.
Such a divergent behaviour would in turn lead to rapid violation
of the $S$-matrix unitarity in the tree approximation (similarly
as in the old Fermi theory) unless there is a special mechanism
suppressing the unwanted terms in physical scattering amplitudes
(for an early reference concerning these matters see e.g. \cite{1}).
It is well known that the present-day standard model (SM) of electroweak
interactions does provide such a mechanism: There are characteristic
subtle cancellations among different Feynman graphs contributing to
a considered $S$-matrix element, so that the tree-level physical
scattering amplitudes are bounded in the high-energy limit
(an example of such a non-trivial divergence cancellation was
first observed by Weinberg \cite{2}). The resulting partial-wave
amplitudes then grow at most logarithmically, and the corresponding
"unitarity bounds" are thus shifted to astronomically high values. What is
even more important, such a good high-energy behaviour of scattering
amplitudes at the tree level ("tree unitarity" \footnote {The technical term
"tree unitarity" means that the $n$-particle $S$-matrix
elements do not grow more rapidly than $E^{4-n}$ in the
$E \to \infty$ limit.}) is in fact a
necessary condition for perturbative renormalizability. (Strictly
speaking, there is no complete rigorous proof of this remarkable
connection, but the usual arguments, based mostly on dispersion
relations -- see e.g. \cite{3} -- indicate that the statement is valid
beyond any reasonable doubt; needless to say, there is also no
known counter-example).

The above-mentioned delicate cancellations of diverging
contributions coming from different Feynman diagrams may appear
"miraculous" in the context of a straightforward calculation in the
"physical" $U$-gauge. In fact, this spectacular phenomenon can be
traced to the original (spontaneously broken) gauge symmetry,
which is completely hidden in the $U$-gauge. The first general
proof of the tree unitarity in spontaneously broken gauge theories
was given by Bell \cite{4}. A discussion of the inverse problem, which
in a sense is even more interesting, followed immediately: In
particular, Llewellyn Smith \cite{5}, and independently Cornwall,
Tiktopoulos and Levin \cite{3} and Joglekar \cite{6} have shown (under
some simple additional constraints) that a theory involving
massive charged vector bosons, which is to satisfy the requirement of the
tree unitarity, must be a non-abelian gauge theory with the Higgs
mechanism realized by means of elementary scalars (for a pedagogical
derivation of the SM along these lines, see e.g. \cite{7}).

At present, a most powerful tool giving clear insight into the
nature of the high-energy behaviour of the scattering amplitudes in SM
(and in other models of this class) is the so-called Equivalence
Theorem (ET) which relates a physical $S$-matrix element,
involving external longitudinally polarized vector bosons, to its
formal "unphysical counterpart" where the longitudinal vector
bosons are replaced by the corresponding (unphysical) Higgs - Goldstone
scalars (unphysical because they actually disappear from the
physical spectrum as a result of the Higgs mechanism).

A statement of this kind has already been mentioned in \cite{3} and
several instructive examples were given by Vayonakis \cite{8}. ET has
subsequently been formulated by Lee, Quigg and Thacker \cite{9} along
with a sketch of the proof in the simplest case of one external
longitudinal vector boson. The first attempt at a general proof is due to
Chanowitz and Gaillard \cite{10}, and somewhat later it was put into a
particularly nice and simple form by Gounaris, K\"ogerler and
Neufeld \cite{11}.
These treatments were then improved by including properly the
relevant renormalization factors arising beyond the tree level;
such a program was started by Yao and Yuan \cite{Yao}, followed
by Bagger and Schmidt \cite{Bagger}, Kilgore \cite{Kilgore}, and
completed by He, Kuang and Li in the series of papers \cite{He}.
Further aspects of ET are still investigated
even in very recent literature,
in particular in connection with the discussion of the mechanism
of electroweak symmetry breaking both within and beyond
SM (cf. e.g. \cite{12}, \cite{13}, \cite{He1995}, \cite{Espriu}).
Some other papers dealing with the subject may be
found under ref. \cite{Willenbrock}.
One should also mention an earlier pedagogical treatment by
Peskin \cite{15} and the monograph \cite{16}
where this topic is also
discussed.

The aim of the present lectures is to provide an introduction to the
Equivalence Theorem, which undoubtedly represents one of the very
remarkable aspects of modern gauge theories. We thus supplement
partly the material of the book \cite{7}, devoted to the theme of
divergence cancellations in scattering amplitudes at high energies
within SM. In order to make these lecture notes rather self-contained,
the conventional formulation of SM is summarized concisely in
the Appendix.

\section{$R$-gauges in the Standard Model}
\vspace{5mm}
\noindent
From the preliminary statement of the Equivalence Theorem mentioned
briefly in the Introduction it is clear that for its formulation
one has to use such a formal description of SM, in which the unphysical
Higgs-Goldstone fields are preserved as
auxiliary variables. This is achieved by using a class of the
so-called renormalizable gauges (or simply $R$-gauges). The
$R$-gauge technique, originally due to 't Hooft \cite{17} was
further developed by Fujikawa, Lee and Sanda \cite{18} and nowadays
it is widely used in practical Feynman diagram calculations in SM.

First, recall how the $U$-gauge is defined (see Appendix).
Fixing the $U$-gauge
means to eliminate completely the would-be (i.e. unphysical)
Goldstone bosons which in a well-known sense are ``natural
partners'' of the massive vector bosons emerging from the Higgs
mechanism. (Setting the three Goldstone fields identically to
zero is feasible since it is formally equivalent to a $SU(2)$
gauge transformation.) A nice feature of the $U$-gauge is that the
corresponding interaction Lagrangian is relatively simple as it
involves only physical fields (for a summary see e.g. Appendix K
in \cite{7}). However, there is a
price to be paid for this convenience: The $U$-gauge propagator of
a massive vector boson ($W$ or $Z$) has the canonical form
\begin{equation}
D_{\mu \nu} (k) = \frac{-g_{\mu \nu} + m^{-2} k_{\mu}
k_{\nu}}{k^{2} - m^{2} + i \varepsilon}
\end{equation}
where $m$ is a corresponding mass. Obviously, it does not
decrease for $k \rightarrow \infty$ like the ``normal''
propagators do, and this in turn leads to a highly divergent
behaviour of Feynman diagrams.

The $R$-gauges were invented in order to tame the severe divergences
of Feynman diagrams in non-abelian gauge theories with
Higgs mechanism. Fixing an $R$-gauge is completely different from
the $U$-gauge case. To define it, one keeps the would-be Goldstone
bosons in the game as auxiliary unphysical fields; one then adds
to the original gauge invariant Lagrangian a non-invariant
``gauge-fixing term'' involving both vector fields and the
unphysical scalars in the manner similar to the familiar ``Fermi
trick'' in QED (which consists in adding a term proportional to
$(\partial \cdot A)^{2}$ to the Maxwell Lagrangian).

Let us now specify the outlined procedure more precisely. The
original complex Higgs doublet is conveniently parametrized as
\begin{eqnarray}
\Phi = \left( \matrix{ & - i w^{+} \cr
& \frac{1}{\sqrt{2}}(v + H + i z) \cr } \right)
\end{eqnarray}
where $w^{+}$ is complex and $H$ and $z$ are real (the constant
$v$ has the usual meaning). Note that by shifting the lower
component of the doublet one gets mass terms for vector bosons
(and for the $H$, of course), so one may define $W_{\mu}^{\pm}$,
$Z_{\mu}$, $A_{\mu}$ instead of $\overrightarrow{A_{\mu}}$,
$B_{\mu}$ as usual (see (A.7), (A.9)).
 The gauge-fixing term is taken to be
\begin{equation}
{\cal{L}}_{g.f.} = - \frac{1}{2 \xi} | \partial^{\mu}
W_{\mu}^{-} - \xi m_{W} w^{-} |^{2} - \frac{1}{2 \xi} |
\partial^{\mu} W_{\mu}^{+} - \xi m_{W} w^{+} |^{2}
\end{equation}
\begin{eqnarray}
\nonumber
\quad
- \frac{1}{2 \eta} ( \partial^{\mu} Z_{\mu} - \eta m_{Z} z )^{2}
- \frac{1}{2 \alpha} ( \partial^{\mu} A_{\mu} )^{2}
\end{eqnarray}
The meaning of the choice (3) is that the quadratic
part of the Lagrangian then becomes diagonal (a mixing of the
$w^{\pm}$ and $W^{\pm}$ etc. is removed). Indeed, by
substituting (2) into the original gauge invariant SM
Lagrangian one gets, in particular
\begin{equation}
(D_{\mu} \Phi)^{\dagger} (D^{\mu} \Phi) = m_{W} ( \partial^{\mu}
w^{-} W_{\mu}^{+} + \partial^{\mu} w^{+} W_{\mu}^{-} ) + m_{Z}
\partial^{\mu} z Z_{\mu} + \ldots
\end{equation}
In (4) we have written explicitly only the
above-mentioned mixing terms involving vector bosons and their
unphysical scalar partners. On the other hand, the gauge-fixing
term (3) also produces some bilinear terms of the
above-mentioned type, namely
\begin{equation}
{\cal{L}}_{g.f.} = m_{W} ( w^{-} \partial^{\mu} W_{\mu}^{+} +
w^{+} \partial^{\mu} W_{\mu}^{-} ) + m_{Z} z \partial^{\mu}
Z_{\mu} + \ldots
\end{equation}
Adding (4) and (5) one gets
\begin{equation}
{\cal{L}}_{Higgs} + {\cal{L}}_{g.f.} = m_{W}
\partial^{\mu} ( w^{-} W_{\mu}^{+} + w^{+} W_{\mu}^{-} ) + m_{Z}
\partial^{\mu} ( z Z_{\mu} ) + \ldots
\end{equation}
i.e. the bilinear terms combine into four-divergences and may
therefore be omitted as we have already indicated above.

Let us now examine the remaining quadratic terms involving
unphysical scalars. From ${\cal{L}}_{Higgs}$ (see (A.28)) one
gets readily kinetic terms for the $w^{\pm}$ and $z$ and
${\cal{L}}_{g.f.}$ yields the corresponding ``mass terms''.
Taken together, these terms amount to
\begin{equation}
{\cal{L}}_{Higgs} + {\cal{L}}_{g.f.} =
\partial^{\mu} w^{-} \partial_{\mu} w^{+} + \frac{1}{2}
\partial^{\mu} z \partial_{\mu} z - \xi m_{W}^{2} w^{-} w^{+} -
\frac{1}{2} \eta m_{Z}^{2} z^{2} + \ldots
\end{equation}
From (7) one may read off the mass parameters
\begin{equation}
m_{w}^{2} = \xi m_{W}^{2}, \quad m_{z}^{2} = \eta m_{Z}^{2}
\end{equation}
These expressions exhibit a dependence of the $w^{\pm}$ and $z$
``masses'' on the (arbitrary) gauge parameters $\xi$, $\eta$;
this reflects the unphysical nature of the would-be Goldstone
bosons which actually disappear from the physical spectrum. On
the other hand, let us emphasize that the mass term for the
physical Higgs boson $H$ comes from the potential (A.30)
upon the shift involved in (2) and thus it is of course
gauge-independent (one has $m_{H}^{2} = 2 \lambda v^{2}$).

Quadratic terms involving vector boson fields descend from
${\cal{L}}_{gauge}$ (kinetic terms --- see
(A.25)), from ${\cal{L}}_{Higgs}$ (mass
terms for $W$ and $Z$) and from ${\cal{L}}_{g.f.}$. One may
summarize them as
\begin{eqnarray}
\nonumber
{\cal{L}}_{gauge} +
{\cal{L}}_{Higgs} + {\cal{L}}_{g.f.} =
-\frac{1}{2} W_{\mu \nu}^{-} W^{+ \mu \nu} - \frac{1}{\xi}
(\partial \cdot W^{-})(\partial \cdot W^{+}) + m_{W}^{2}
W_{\mu}^{-} W^{+ \mu}
\end{eqnarray}
\begin{equation}
\qquad
- \frac{1}{4} Z_{\mu \nu} Z^{\mu \nu} - \frac{1}{2 \eta}
(\partial \cdot Z)^{2} + \frac{1}{2} m_{Z}^{2} Z_{\mu} Z^{\mu}
\end{equation}
\begin{eqnarray}
\nonumber
\qquad
-\frac{1}{4} A_{\mu \nu} A^{\mu \nu} - \frac{1}{2 \alpha} (
\partial \cdot A)^{2} + \ldots
\end{eqnarray}
In quantized theory, propagators of vector bosons are obtained
by inverting the quadratic form in (9) (for a standard
technique of doing so, see e.g. Appendix H in \cite{7}).
 The result is
\begin{eqnarray}
\nonumber
D_{\mu \nu}^{(W)} (k) = [ - g_{\mu \nu} + ( 1 - \xi ) ( k^{2} -
\xi m_{W}^{2})^{-1} k_{\mu} k_{\nu} ] \frac{1}{k^{2} - m_{W}^{2}
+ i \varepsilon}
\end{eqnarray}
\begin{equation}
\label{eq91}
D_{\mu \nu}^{(Z)} (k) = [ - g_{\mu \nu} + ( 1 - \eta ) ( k^{2} -
\eta m_{W}^{2})^{-1} k_{\mu} k_{\nu} ] \frac{1}{k^{2} - m_{Z}^{2}
+ i \varepsilon}
\end{equation}
\begin{eqnarray}
\nonumber
D_{\mu \nu}^{(A)} (k) = [ - g_{\mu \nu} + ( 1 - \alpha ) (
k^{2})^{-1}  k_{\mu} k_{\nu} ] \frac{1}{k^{2} + i \varepsilon}
\end{eqnarray}

The expressions (10) are seen to behave like $k^{-2}$
for $k \rightarrow \infty$ and this explains the term
``renormalizable gauges''. Let us stress again that such a
decent ultraviolet behaviour of the massive vector boson
propagators has been achieved at the price of introducing
unphysical degrees of freedom --- the ``Goldstone scalar
ghosts'' $w^{\pm}$ and $z$. Let us also add that for a general
$R$-gauge (3) the propagators of $w^{\pm}$ and $z$ are,
in view of (7)
\begin{equation}
D^{(w)}(k) = \frac{1}{k^{2} - \xi m_{W}^{2} + i \varepsilon}
\end{equation}
\begin{eqnarray}
\nonumber
D^{(z)}(k) = \frac{1}{k^{2} - \eta m_{Z}^{2} + i \varepsilon}
\end{eqnarray}
From (10) it is obvious that for practical calculations,
the most convenient choice of gauge corresponds to $\xi = \eta =
\alpha = 1$ as the vector boson propagators are then diagonal;
this is the familiar 't Hooft -- Feynman gauge. Note that in
this gauge the (unphysical) mass parameters of the scalars
$w^{\pm}$, $z$ are equal to the masses of their vector boson
counterparts $W^{\pm}$, $Z$. One may also notice that taking the
limit $\xi \rightarrow \infty$, $\eta \rightarrow \infty$ in
(10) one recovers the $U$-gauge propagators of $W$ and $Z$
(cf. (11)). Furthermore, the expressions (11) are
seen to vanish identically in such a limit (the $w^{\pm}$, $z$
become ``infinitely heavy''); this is of course gratifying since
the $w^{\pm}$ and $z$ should be absent in the $U$-gauge (by
definition).

It is clear that the SM interaction Lagrangian in an $R$-gauge will
contain many additional contributions in comparison with the
$U$-gauge case since now one must also consider terms involving
the unphysical scalars. A complete catalogue of the $R$-gauge
interaction vertices may be found in many places (see e.g.
\cite{19}, \cite{20}).
 Here we will restrict ourselves only to some instructive examples.

First, when (2) is used in
${\cal{L}}_{Higgs}$, the expression $(D^{\mu}
\Phi)^{\dagger} (D_{\mu} \Phi)$ yields, among other things, a
trilinear interaction involving two scalars $w^{\pm}$ and a
(neutral) vector boson, which reads
\begin{equation}
\label{eq93}
{\cal{L}}_{w^{-} w^{+} V} = i e w^{-}
{\overleftrightarrow{\partial^{\mu}}} w^{+} A_{\mu} + i
\frac{g}{\cos \vartheta_{W}} ( \frac{1}{2} - \sin^{2}
\vartheta_{W} ) w^{-} {\overleftrightarrow{\partial^{\mu}}}
w^{+} Z_{\mu}
\end{equation}
In this context let us remark that many conceivable types of
$R$-gauge interaction terms may be formally deduced from $U$-gauge
vertices by replacing one or more vector boson lines by the
corresponding unphysical Goldstone bosons. The expression
(12) is an explicit example of such an interaction term.

Second, considering the ${\cal{L}}_{Yukawa}$ for
a lepton $l$ (see (A.41)) one gets, using (2)
\begin{equation}
{\cal{L}}_{Yukawa} = i \frac{g}{2 \sqrt{2}}
\frac{m_{l}}{m_{W}} {\overline{\nu}} (1 + \gamma_{5}) l w^{+} +
{\rm h.c.}
\end{equation}
\begin{eqnarray}
\nonumber
\qquad
- i \frac{g}{2} \frac{m_{l}}{m_{W}} {\overline{l}} \gamma_{5} l
z - \frac{g}{2} \frac{m_{l}}{m_{W}} {\overline{l}} l H
\end{eqnarray}
(the standard lepton--Higgs interaction is of course recovered
in (13) as expected).

Third, there are new scalar self-interactions descending from
the ``potential'' $V(\Phi)$ in
${\cal{L}}_{Higgs}$. One has
\begin{equation}
\label{eq95}
{\cal{L}}^{\mbox{\sixrm{(scalar)}}}_{\mbox{\sixrm{int.}}} = -
\lambda v H ( 2 w^{-} w^{+} + z^{2} + H^{2}) - \frac{1}{4}
\lambda ( 2 w^{-} w^{+} + z^{2} + H^{2})^{2}
\end{equation}
Again, (14) incorporates also the Higgs boson
self-interactions known from the $U$-gauge formulation.

All this, however, is not the whole story yet. The standard model
is a non-abelian gauge theory and when it is quantized in an
$R$-gauge (3), one has to introduce another set of
unphysical fields, namely the Faddeev--Popov (FP) ghosts (which
do not occur in the $U$-gauge). Note that an essential reason for
invoking the FP ghosts in a gauge like (3) is that
otherwise the $S$-matrix would not be unitary at one-loop level
(see e.g. \cite{20}, \cite{21}).
Thus, a complete relevant Lagrangian in the considered $R$-gauge
reads, schematically
\begin{equation}
\label{eq96}
{\cal{L}}^{(R-gauge)}_{SM} =
{\cal{L}}_{g.inv.} +  {\cal{L}}_{g.f.} +
{\cal{L}}_{FP}
\end{equation}
The FP term can be derived most efficiently by means of the path-integral
techniques (see \cite{19}). A detailed form of the ${\cal{L}}_{FP}$ will
not be needed in what follows, so we do not reproduce it here.
For the purpose of later references we will only summarize briefly
some essential features of the quantum $R$-gauge SM Lagrangian
(15).

First let us introduce a convenient shorthand notation
for the gauge-fixing (GF) functions occurring in
(3), namely
\begin{equation}
F_a [V, \varphi] = \cases {\partial^{\mu} W_{\mu}^{\pm} - \xi m_W w^{\pm}\cr
\partial^{\mu} Z_{\mu} - \eta m_Z z\cr
\partial^{\mu} A_{\mu}\cr}
\end{equation}
(on the l.h.s. of (16), the index $a$ labels the four SM vector
fields denoted collectively by $V$, and $\varphi$ stands for
the unphysical Higgs-Goldstone fields).

The FP term in (15) involves four ghost fields $c_a$ (associated
with the four gauge bosons) and the corresponding conjugate
(antighost) variables $\bar {c}_a$. The structure of ${\cal {L}}_{FP}$
is determined by the gauge variation of the GF functions $F_a$
(see \cite{19}). FP ghosts represent (unphysical) Lorentz scalars
obeying Fermi statistics and enter Feynman diagrams only via closed
loops. Let us remark that the corresponding mass-squared parameters are
$\xi m_W^2$ for $c_{\pm}$, $\eta m^2_Z$ for $c_Z$ and 0 for
$c_{\gamma}$. Further, the interaction terms contained in ${\cal {L}}_{FP}$
are such that a pair of FP ghosts (being not both neutral) is
coupled to another field which may be a vector $(W^{\pm}, Z$ or
$\gamma$), the Higgs boson $H$, or an unphysical Goldstone scalar
($w^{\pm}$ or $z$).

The most remarkable property of the full SM Lagrangian (15) is a
peculiar {\it global} symmetry discovered by Becchi, Rouet and Stora
\cite{22} (and independently by Tyutin \cite{23}) which represents, in a sense,
a "remnant" of the original classical gauge symmetry, broken by the
gauge-fixing procedure, i.e. by including
the ${\cal {L}}_{g.f.}$ and the associated term ${\cal {L}}_{FP}$. Such
a "residual" symmetry is in fact a general feature of quantized gauge
theories (for a review, see \cite{24}) and nowadays it is an issue
discussed in most textbooks on modern field theory (see e.g.
\cite{19}, \cite{20}, \cite{21}).
In the present context, the BRS transformations may be
written (rather schematically) as
\begin{eqnarray}
\delta V_a^{\mu} &=& \theta D_{ab}^{\mu} c_b\nonumber\\
\delta \Psi &=& -i \theta T_a  c_a \Psi\\
\delta \bar{c}_a &=& - \theta {1 \over \xi} F_a\nonumber\\
\delta c_a &=& - \theta {1 \over 2} f_{abd} c_b c_d\nonumber
\end{eqnarray}
where $D^{\mu}$ is the relevant covariant derivative (in the adjoint
representation), $\Psi$ is a generic symbol for the matter fields,
$T_a$ is a gauge group generator, $f_{abd}$ denotes a corresponding
structure constant, and $F_a$ is given by (16). For simplicity,
all gauge-fixing parameters are denoted by $\xi$ and we have also
suppressed the coupling constants. The parameter $\theta$ is a constant
{\it anticommuting} (Grassmann) real number; it means that
$\theta^2 = 0$ and $\theta$ is thus effectively infinitesimal.

Now it is well known (see e.g. \cite{20}, \cite{21}) that by using the BRS
symmetry one can recover the Ward-Takahashi (Slavnov-Taylor)
 identities of a quantized
gauge theory. Thus, the global BRS symmetry describes concisely
the contents of the local gauge invariance at quantum level.
Quantization of non-abelian gauge theories is usually implemented in the
path-integral formalism. However, as we shall see later, in some
situations it may be useful to have at hand also a covariant
canonical operator method. Such a quantization procedure (which,
roughly speaking, is a generalization of the well-known Gupta-Bleuler
method in QED) was indeed invented by Kugo and Ojima \cite{25} (see
also \cite{26}, \cite{27} and references therein). In their approach, the notion
of BRS symmetry plays a crucial role. The main breakthrough of
\cite{25} consists in a successful generalization of the
Gupta-Bleuler subsidiary condition, characterizing a subspace
of physical states. Let us recall that the GB condition may be
written as
\begin{equation}
\partial^{\mu} A_{\mu}^{(-)} (x) \vert \psi_{\rm phys.} > = 0
\end{equation}
where $A_{\mu}^{(-)} (x)$ denotes the annihilation part of quantized
electromagnetic potential. According to \cite{25}, a correct
generalization of (18) to the non-abelian case is  astonishingly
simple, yet highly non-trivial; it reads
\begin{equation}
Q_{BRS} \vert \psi_{\rm phys.} > = 0
\end{equation}
where $Q_{BRS}$ is an operator version of the conserved Noether
charge corresponding to BRS symmetry (note that for an abelian gauge field
it can be shown that (19) implies (18), so (19) may indeed be
viewed as a natural extension of (18) in a general case). The
canonical (anti)commutation relations set up in \cite{25} ensure that
$Q_{BRS}$ is the generator of transformations (17) for the
corresponding field operators (let us emphasize that the FP ghosts
satisfy anticommutation relations). An example of such
an operator relation, which will be useful in subsequent discussion, is
provided by the anticommutator
\begin{equation}
\{ Q_{BRS} , \bar{c_a}\} = -{1 \over \xi}F_a
\end{equation}
(note that (20) corresponds to the BRS transformation of the
antighost field in (17)). Let us also add that the BRS charge has
another remarkable property: It is {\it nilpotent}, i.e.
\begin{equation}
Q_{BRS}^2 = 0
\end{equation}
(it follows from (21) that BRS transformations of state vectors
and operators are effectively infinitesimal).

In the section 4 we will invoke only some basic ingredients of
the Kugo-Ojima canonical operator quantization scheme mentioned
above, at the level necessary for understanding a basic idea of
the proof of Equivalence Theorem for longitudinal vector bosons.

\vspace{0,5cm}
\noindent
{\bf{Exercise:}} {\it{At the tree level prove that the
scattering amplitudes for $e^{+} e^{-} \longrightarrow \mu^{+}
\mu^{-}$ in a $R$-gauge and in $U$-gauge are equal.}}

\section{Equivalence Theorem -- examples}

\noindent
A familiar part of the physical ``folklore'' is a rather vague
but frequently used statement concerning the Higgs mechanism,
which may be paraphrased roughly as follows:

The would-be
Goldstone bosons are ``eaten'' by the gauge fields which become
massive and the massive vector bosons may have --- in contrast
to massless ones --- also longitudinal polarizations. Thus, in a
sense, the unphysical Goldstone scalars become
the longitudinal components of massive vector bosons.

It is
remarkable that such an intuitive statement may be given a more
precise meaning on the level of $S$-matrix elements.
Indeed, this is the contents of the Equivalence Theorem \cite{8} --
\cite{He},
stating that in high-energy limit, the $S$-matrix element for a
process with external longitudinally polarized vector bosons is
equal, up to a constant factor, to a matrix element (calculated
within $R$-gauge) in which longitudinal vector bosons are replaced
by the corresponding unphysical Higgs-Goldstone scalars.
We will formulate the theorem explicitly later in this section;
now we are going to give two instructive
examples of how such an equivalence works ``in practice''.

As a first example let us consider the decay of a very heavy
Higgs boson ($m_{H} \gg 2 m_{W}$) into a pair of longitudinally
polarized vector bosons $W^{\pm}$. In lowest order, the process
is described by the diagram

\newpage
\epsfig{file=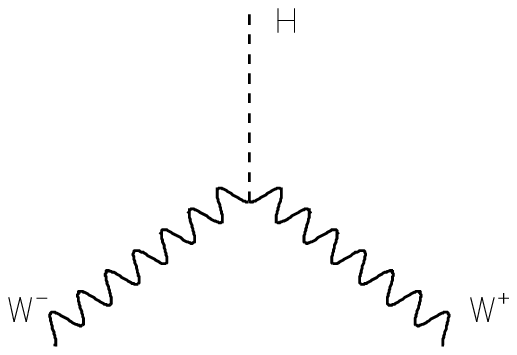}

\noindent
and the corresponding Lorentz invariant matrix element may be
written as (cf. A.38))
\begin{equation}
{\cal{M}} ( H \rightarrow W_{L}^{-} W_{L}^{+}) = g m_{W}
\varepsilon_{L}^{\mu} (k) \varepsilon_{L \mu} (p)
\end{equation}
\begin{eqnarray}
\nonumber
\qquad
= g m_{W} ( m_{W}^{-1} k^{\mu} + \Delta^{\mu}(k) )(m_{W}^{-1}
p_{\mu} + \Delta_{\mu}(p))
\end{eqnarray}
In (22) we have used the well-known high-energy
decomposition of longitudinal polarization vectors (for details
see e.g. Appendix H in \cite{7}); note that
the quantities $\Delta^{\mu}(k)$, $\Delta^{\mu}(p)$ are of an
order of ${\cal{O}}(m_{W}/E_{W})$ where of course $E_{W} =
\frac{1}{2} m_{H}$, i.e. $\Delta^{\mu} \ll 1$ in our case. Thus,
from (22) one gets, after some simple manipulations
\begin{equation}
{\cal{M}}(H \rightarrow W_{L}^{-} W_{L}^{+}) = \frac{g}{2}
\frac{m_{H}^{2}}{m_{W}} ( 1 + {\cal{O}}(m_{W}^{2}/E_{W}^{2}))
\end{equation}
On the other hand, for the unphysical process $H \rightarrow
w^{-} w^{+}$ described formally within an $R$-gauge by the
lowest-order diagram

\epsfig{file=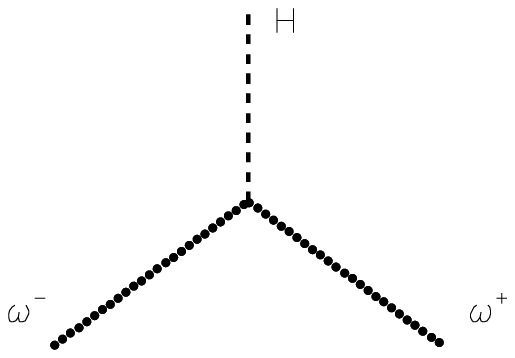}

\noindent
one gets the matrix element (see (14))
\begin{equation}
{\cal{M}}(H \rightarrow w^{-} w^{+}) = - \frac{g}{2}
\frac{M_{H}^{2}}{m_{W}}
\end{equation}
Comparing (23) and (24) one may observe that
these two matrix elements are indeed equal, up to a correction
suppressed by a factor of $m_{W}^{2}/E_{W}^{2}$ and up to a
minus sign.

As another, less trivial example let us consider the process
$e^{+} e^{-} \rightarrow W_{L}^{-} W_{L}^{+}$. For simplicity,
we will assume $m_{e} = 0$. The lowest-order (i.e. tree-level)
scattering amplitude for such a process is a sum of several
Feynman diagrams (as an exercise, draw these diagrams in $U$-gauge and in $R$-gauge
resp.) and one gets, after a rather long calculation
\begin{equation}
\hspace{-3cm}
{\cal{M}}(e^{+} e^{-} \rightarrow W_{L}^{-} W_{L}^{+}) =
\frac{1}{s} {\overline{v}}(k_{2}) (\rlap/{p}_{1} -
\rlap/{p}_{2}) u(k_{1}) \times
\end{equation}
\begin{eqnarray}
\nonumber
\times
[ e^{2} + g^{2} ( -\frac{1}{2} +
\sin^{2} \vartheta_{W} )( \frac{m_{Z}^{2}}{2 m_{W}^{2}} - 1)] +
{\cal{O}}(m_{W}^{2}/s)
\end{eqnarray}
where $k_{1}$, $k_{2}$, $p_{1}$ and $p_{2}$ are consecutively
the four-momenta of $e^{-}$, $e^{+}$, $W^{-}$ and $W^{+}$ and
$s$ is the usual Mandelstam invariant. On the other hand, for
the unphysical process $e^{+} e^{-} \rightarrow w^{-} w^{+}$ one
gets (e.g. in the 't Hooft -- Feynman gauge)
\begin{equation}
\hspace{-3cm}
{\cal{M}}(e^{+} e^{-} \rightarrow w^{-} w^{+}) =
- \frac{1}{s} {\overline{v}}(k_{2}) (\rlap/{p}_{1} -
\rlap/{p}_{2}) u(k_{1}) \times
\end{equation}
\begin{eqnarray}
\nonumber
\times
[ e^{2} + (\frac{g}{\cos
\vartheta_{W}})^{2}  ( \frac{1}{2} -
\sin^{2} \vartheta_{W} )^{2}] +
{\cal{O}}(m_{W}^{2}/s)
\end{eqnarray}
Note that the matrix element (26) is again obtained by
summing several tree-level Feynman diagrams (which graphs do
actually contribute ?); the rules for the relevant vertices
follow from (12) and (13). Now, using in
(26) the relation $m_{W}/m_{Z} = \cos \vartheta_{W}$,
the expressions (25) and (26) are seen to
coincide, up to a sign and barring the corrections suppressed by
$m_{W}^{2}/s \ll 1$.

The above two examples illustrate the general theorem
 mentioned earlier in this section, namely\\

 \noindent
 {\bf Equivalence Theorem:} Let us consider a process involving,
 apart from other physical particles, a certain number of
 longitudinally polarized vector bosons $V_L$ (i.e. $W_L^{\pm}$
 and/or $Z_L$), with $n_1$ of them being in the initial state and
 $n_2$ in the final state. Let $E_V$ denote generically the
 vector-boson energies; for $E_V \gg m_V$ one then has
\begin{eqnarray}
&&{\cal{M}}_{fi}(V_{L}(i_1), \dots, V_{L} (i_{n_1}), A \to
V_L (f_1), \dots, V_L (f_{n_2}), B) =\nonumber\\
&=& {\cal{M}}_{fi}
(\varphi (i_1), \dots, \varphi (i_{n_1}), A \to \varphi (f_1), \dots,
\varphi (f_{n_2}), B) \times\nonumber\\
&\times& i^{n_1} (-i)^{n_2} C [1 + O (m_V/E_V)]
\end{eqnarray}
where the $\varphi$'s stand for the unphysical Higgs-Goldstone
scalar counterparts of the $V'_L$s, the $A, B$ symbolize all
other incoming and outgoing particles and the factor $C$ is in general
a constant independent of the energies which is due to renormalization
effects ($C$ is a product of factors associated with individual
external Higgs-Goldstone fields, $C = 1$ at the tree level). $\spadesuit$

Note that the phase factor contained in (27) is indeed recovered
in our two previous examples (where $n_1 = 0, n_2 = 2$). The idea
of a general proof of ET is the subject of the next section
(except the discussion of the $C$, for which the reader is
referred to the original literature \cite{Yao} -- \cite{He}).

\noindent
{\bf Exercise:} {\it Verify the validity of ET (27) at the tree
level in the case
of decay $l \to W + \nu_l$, where $l$ is a hypothetical heavy lepton
($m_l \gg m_W$) with usual SM couplings.}

\section {Idea of ET proof -- a basic Ward identity}

While the explicit examples given in preceding section
clearly support the validity of ET (27), one would like
to know what is actually the true origin of such a
remarkable statement. In this section we will outline
briefly the idea of a formal proof of ET, following
essentially the work of Gounaris et al. \cite{11}.

Intuitively, a connection between longitudinal vector
bosons and unphysical Higgs-Goldstone scalars is
suggested by the form of the GF functions (16),
and one may expect that ET should follow formally
from an appropriate Ward identity reflecting the
gauge invariance of the considered electroweak
theory. This is indeed the case. The foundation for
a formal proof of ET is provided by the identity
\begin{equation}
< B \vert T [
F_{a_1} (x_1) \dots F_{a_n} (x_n) ] \vert A >_{con} = 0
\end{equation}
where $F_a (x)$ is given by (16) and $\vert A >, \; \vert B >$
are {\it physical} in- and out-states (i.e. they satisfy the
condition (19)). The subscript at the matrix element
denotes its "connected part": It means that the terms, which
correspond to factorization of a vacuum-to-vacuum matrix
element involving the $F'_a$s, are discarded (it is shown in \cite{11}
that such terms have contact character - they are proportional to
products of the delta-functions $\delta^4 (x_j - x_k)$).
First we are going to discuss the origin of (28) and its
connection with ET will be explained later. We are using the
framework of canonical operator quantization of non-abelian
gauge theories established in \cite{25}, \cite{26}, \cite{27} (note that the
treatment of ref. \cite{11} does not rely explicitly on the
operator formalism).

One may start with the following observation: For a matrix element
of a string of local operators between physical states (characterized
by (19)) one gets the identity
\begin{equation}
0 = \sum^{n}_{k-1} < B \vert T [ O_1 (x_1) \dots O_{k-1} (x_{k-1})
\delta_{BRS} O_k (x_k) O_{k+1} (x_{k+1}) \dots O_n (x_n) ] \vert A >
\end{equation}
where the (infinitesimal) BRS transformation of an operator
$O_k$ is given by an (anti)commutator with the generator
$Q_{BRS}$ (cf. the discussion around eq. (20)). Such a relation is
obtained immediately, if one considers an (anti)commutator of
$Q_{BRS}$ with the relevant operator product, sandwiched
between the physical states: On the one hand, this obviously
vanishes owing to (19); on the other hand, working it out one
gets the sum on the r.h.s. of (29). (Note that since (29)
emerges as a consequence of the BRS symmetry of the underlying theory,
it represents the prototype of a general Ward identity.) The
identity (28) can now be proved by utilizing a set of the
relations (29) with operators $O_k (x)$ being either the
antighost fields $\bar{c}_a (x)$ or the GF functions $F_a (x)$.
Starting with $n = 1$, one may choose $O_1 (x) = \bar{c}_a (x)$
in (29); then using (20) it follows immediately
\begin{equation}
< B \vert F_a (x) \vert A > = 0
\end{equation}
In fact, it is not difficult to realize that this already
proves (28) for $n = 1$ (in (30) there cannot be any "disconnected"
term of the above-mentioned type -- this is obvious e.g. from the
explicit form of the $F_a$ as given by (16)). For $n = 2$, one
chooses
\begin{equation}
O_1 (x_1) = F_{a_1} (x_1), \qquad O_2 (x_2)  = \bar{c_{a_2}} (x_2)
\end{equation}
in (29). In this case, one must invoke an additional fact
about BRS symmetry which has not been mentioned so far, namely
that the BRS transform of a gauge-fixing function is
proportional to the equation of motion of the associated
antighost (see e.g. \cite{24}). Equipped with this knowledge, and
using eq. (20) as well, one arrives at the identity (28) with
$n = 2$ (after discarding a disconnected contact term, proportional to
$\delta^4 (x_1 - x_2))$. One may proceed further in this way
following (31), and complete the proof by induction. Here we have only
sketched its basic idea; for more technical details the reader
is referred to \cite{11}. Let us add that a virtue of the operator formalism
in the present context consists mainly in the compact
characterization of physical states by means of the condition (19).

We will now show how ET follows from the identity (28). To
demonstrate a basic idea of the proof, we are going to discuss
in detail first the simplest case involving a single longitudinal
vector boson. Later on we will indicate, by means of a particular
example, how the procedure can be generalized. Thus, let us
consider a process $A \to B + V_a$, where $V_a$ stands for
$W^{\pm}$ or $Z$ and $A, B$ denote other physical particles
(including possibly other vector bosons as well). The
corresponding amplitude, i.e. an $S$-matrix element, can be
expressed by means of an appropriate reduction formula (obviously,
such a representation is useful in view of an envisaged
application of (28)). For a "truncated matrix element" of the
indicated process one may then write, schematically
\begin{equation}
{\cal{M}}_{\mu} (A \to B + V_a (p, \lambda)) =
{\rm FT}  \{ L_{\mu \nu} < B \vert V^{\nu}_a (x) \vert A >\}
\end{equation}
Here "truncation" means removing the polarization vector
of the $V_a$ from the full matrix element ${\cal{M}}$,
expressed as ${\cal{M}} = {\cal{M}}_{\mu} \varepsilon^{\mu}
(p, \lambda)$. The symbol FT in (32) denotes Fourier
transformation and $L_{\mu \nu}$ is the linear
differential operator which appears in the equation
of motion for $V_a$; it reads
\begin{equation}
L_{\mu \nu} = (\Box + m^2_a) g_{\mu \nu} + ({1 \over \xi} - 1)
\partial_{\mu} \partial_{\nu}
\end{equation}
 (note that (33) corresponds to the quadratic part of
 the Lagrangian (9)). Let us also remark that we suppress
 systematically factors related to normalization of
 one-particle states, or those which are due to conventions
 adopted in defining the considered matrix elements. This is
 justified since our result will be a simple linear relation between
 two matrix elements (with a similar structure) in which all
 such extra factors cancel. From (32) one gets further
\begin{equation}
i p^{\mu} {\cal M}_{\mu} = \mbox{FT} \left\{ \partial^{\mu}
L_{\mu \nu} \left< B \left| V_{a}^{\nu} \left( x \right) \right|
A \right> \right\}
\end{equation}

Taking into account (33), this becomes
\begin{equation}
i p^{\mu} {\cal M}_{\mu} = \mbox{FT} \left\{ \left( m_{a}^{2} +
\frac{1}{\xi} \Box \right) \left< B \left| \partial_{\nu}
V_{a}^{\nu} \left( x \right) \right| A \right> \right\}
\end{equation}
According to (16), $\partial_{\nu} V_{a}^{\nu} \left( x \right)$
may be recast as $F_{a} + \xi m_{a} \varphi_{a}$, so in the r.h.s.
of (35) one may employ subsequently the identity (28) with $n=1$
(i.e. eq.(30)). One thus gets
\begin{equation}
i \frac{p^{\mu}}{m_{a}} {\cal M}_{\mu} = \mbox{FT} \left\{ \left(
\Box + \xi m_{a}^{2} \right) \left< B \left| \varphi_{a} \left( x
\right) \right| A \right> \right\}
\end{equation}
However, the expression on the r.h.s. of (36) gives (via a
reduction formula) just the amplitude of the process $A
\longrightarrow B + \varphi_{a} \left( p \right)$. Thus, we arrive
at the identity
\begin{equation}
i \frac{p^{\mu}}{m_{a}} {\cal M}_{\mu} \left( A \longrightarrow B
+ V_{a} \left( p, \lambda\right) \right) = {\cal M} \left( A
\longrightarrow B + \varphi_{a} \left( p \right) \right)
\end{equation}
It should be noted that for a corresponding process with an
{\it incoming} vector boson (carrying four-momentum $p$) one would
obtain an analogous relation with opposite sign in the
left-hand side, i.e. with $i \rightarrow -i$. Let us emphasize
that up to now the polarization of the vector boson was
completely arbitrary, i.e. (37) is valid for any $\lambda$.
Consider now the case $\lambda=L$ and assume also that the states
$A$, $B$ do not contain any other longitudinal vector boson. The
vector of longitudinal polarization can be decomposed as
\begin{equation}
\varepsilon^{\mu}_{L} \left( p \right) = \frac{1}{m_{a}} p^{\mu}
+ \Delta^{\mu} \left( p \right)
\end{equation}
(c.f. (22) etc.) where the $\Delta^{\mu} \left( p \right)$
behaves as ${\cal O} \left( m_{a}/E \right)$ for $E \gg m_{a}$.
Taking into account (38), one then gets immediately
\begin{equation}
{\cal M}_{\mu} \varepsilon_{L}^{\mu} \left( p \right) =
\frac{1}{m_{a}} p^{\mu} {\cal M}_{\mu} + {\cal O} \left(
\frac{m_{a}}{E} \right)
\end{equation}
in the high-energy limit; using (37), it can be further rewritten
as
\begin{equation}
{\cal M} \left( A \longrightarrow B + V_{a} \left( p, \lambda = L
\right) \right) = - i {\cal M} \left( A \longrightarrow B +
\varphi_{a} \left( p \right) \right) \left[ 1 + {\cal O} \left(
\frac{m_{a}}{E} \right) \right]
\end{equation}
Note that in writing (39) and (40) we have also tacitly assumed
that both ${\cal M}_{\mu}$ and ${\cal M} \left( A \longrightarrow
B + \varphi_{a} \right)$ are bounded in the high-energy limit; this
is justified as these quantities do not contain any source of a
``bad'' high-energy behaviour. The relation (40) is just the
statement of the ``naive'' ET for a single longitudinal vector boson in the
final state (cf. (27) with $n_{1} = 0$, $n_{2} = 1$ and $C = 1$). Of course,
for a process with one longitudinal vector boson in the initial
state one would obtain an analogous relation, with the $-i$ being
replaced by $i$ (cf. the remark following eq.(37)). The term
``naive'' refers to the fact that throughout our discussion we
have neglected the relevant renormalization effects which may in
general lead to the finite modification factors incorporated in
eq.~(27). A detailed discussion of a such an issue would go
beyond the scope of this introductory review and the interested
reader is referred to the original papers \cite{Yao} --
\cite{He}. Let us also remark that, according to the analysis
performed in \cite{He}, the modification factor $C$ in
(27) can be made equal to unity in a suitable renormalization
scheme. Anyway, even in the general case with $C \neq 1$
a finite multiplicative constant does not bear on the
issue of the asymptotic behaviour of scattering amplitudes in SM
for energies much larger than any mass involved in the theory.

Having demonstrated the basic idea of ET proof, we are not going
to discuss the general case in detail. Instead, we will only
indicate how one could proceed further: Next we are going to
formulate the relevant generalization of the identity (37) and
show subsequently how it can be employed, by analysing a
particular example. To this end, let us first introduce a useful
notation due to Chanowitz and Gaillard \cite{10}. Following \cite{10}, one
may define five-component objects $D_{M}^{a} \left( a \right)$
and ${\cal M}_{M}$ with $M = 0,1,2,3,4$ as
\begin{eqnarray}
\nonumber
D_{M}^{a} \left( p \right) &=& \left( -ip_{\mu}, m_{a} \right) \\
{\cal M}_{M} &=& \left( {\cal M}_{\mu}, {\cal M}_{4} \right)
\end{eqnarray}
where ${\cal M}_{\mu}$ has the usual meaning and ${\cal M}_{4} =
{\cal M} \left( A \longrightarrow B + \varphi_{a} \right)$; we have
thus introduced formally the (trivial) ``truncation'' of a matrix
element for emission of a scalar $\varphi_{a}$. Note that we do not
distinguish between ${\cal M}_{4}$ and ${\cal M}^{4}$ etc. In
terms of the symbols (41), the identity (37) may be written
compactly as
\begin{equation}
D_{M}^{a} \left( p \right) {\cal M} ^{M} = 0
\end{equation}
The generalization of eq.(37) (or, equivalently, (42)) for matrix
elements involving an arbitrary number of vector bosons and/or
unphysical Higgs-Goldstone scalars corresponds to a general $n$
in the basic Ward identity (28). For simplicity, let us restrict
ourselves to the case where all vector bosons are outgoing. The
relevant identity obtained first in \cite{10} reads
\begin{equation}
D_{M_{1}}^{a_{1}} \left( p_{1} \right) \cdots D_{M_{n}}^{a_{n}}
\left( p_{n} \right) {\cal M}^{M_{1} \dots M_{n}} = 0
\end{equation}
where ${\cal M}^{M_{1} \dots M_{n}}$ denotes a (partially)
truncated matrix element, in the sense specified above. By
``partial truncation'' we mean that such a matrix element may
still incorporate polarization vectors of additional vector
bosons (contained in the physical states $A$, $B$ in (28)). Note
also that in the most general case involving both outgoing and
incoming vector bosons, the extension of eq.(43) includes also
the conjugate ``five-vectors'' $\widetilde{D}_{M}$ given by
\begin{equation}
\widetilde{D}_{M}^{a} \left( p \right) = \left( i p_{\mu}, m_{a}
\right)
\end{equation}
which correspond just to the incoming vector bosons and
unphysical scalars (cf. the remark following eq.(37)).

Chanowitz and Gaillard \cite{10} employed the set of identities (43) to
accomplish a general proof of ET. Here we will only illustrate
how eq.(43) can be utilized in such an analysis, by considering a
particular example, namely the process $e^{+} e^{-}
\longrightarrow W^{-}_{L} W^{+}_{L}$ (which we have discussed at
an elementary level in preceding section --- cf.(26), (26)). To
begin with, let us introduce the truncated matrix elements
associated with the considered process; these are defined by
\begin{eqnarray}
\nonumber
{\cal M} \left( e^{-} e^{+} \longrightarrow W^{-} W^{+} \right)
&=& {\cal M}_{\mu \nu} \varepsilon^{\mu} \left( p_{1} \right)
\varepsilon^{\nu} \left( p_{2} \right)
\\
\nonumber
{\cal M} \left( e^{-} e^{+} \longrightarrow w^{-} W^{+} \right)
&=& {\cal M}_{4 \nu} \varepsilon^{\nu} \left( p_{2} \right)
\\
\nonumber
{\cal M} \left( e^{-} e^{+} \longrightarrow W^{-} w^{+} \right)
&=& {\cal M}_{\mu 4} \varepsilon^{\mu} \left( p_{1} \right)
\\
{\cal M} \left( e^{-} e^{+} \longrightarrow w^{-} w^{+} \right)
&=& {\cal M}_{4 4}
\end{eqnarray}
(we denote the relevant four-momenta as in (25), (26)). In the
subsequent discussion we will employ the identities (43) with
$n=1$ and $n=2$, namely
\begin{eqnarray}
\nonumber
D^{M} \left( p_{1} \right) {\cal M}_{M \nu} \varepsilon^{\nu}
\left( p_{2} \right) &=& 0
\\
D^{N} \left( p_{2} \right) {\cal M}_{\mu N} \varepsilon^{\mu}
\left( p_{1} \right) &=& 0
\end{eqnarray}
and
\begin{equation}
D^{M} \left( p_{1} \right) D^{N} \left( p_{2} \right) {\cal
M}_{MN} = 0
\end{equation}
(to simplify the notation, we have suppressed here the labels
$a_{j}$). Using the definition (41), the relations (46) and (47)
may be worked out as
\begin{eqnarray}
\nonumber
\left( - i p_{1}^{\mu} {\cal M}_{\mu \nu} + m_{W} {\cal M}_{4\nu}
\right) \varepsilon^{\nu} \left( p_{2} \right) &=& 0
\\
\left( - i p_{2}^{\nu} {\cal M}_{\mu \nu} + m_{W} {\cal M}_{\mu 4}
\right) \varepsilon^{\mu} \left( p_{1} \right) &=& 0
\end{eqnarray}
and
\begin{equation}
- p_{1}^{\mu} p_{2}^{\nu} {\cal M}_{\mu \nu} - i p_{1}^{\mu}
m_{W} {\cal M}_{\mu 4} - i p_{2}^{\nu} m_{W} {\cal M}_{4 \nu} +
m_{W}^{2} {\cal M}_{44} = 0
\end{equation}
(up to now, vector boson polarizations may be arbitrary). Let us
now show that ET is valid for the considered process. Denoting
the amplitude for $e^{-} e^{+} \longrightarrow W^{-}_{L}
W^{+}_{L}$ simply by ${\cal M}_{LL}$ and employing the
decomposition (38) for longitudinal polarization vectors, we get
first
\begin{eqnarray}
{\cal M}_{LL} &=& {\cal M}_{\mu \nu} \varepsilon_{L}^{\mu} \left(
p_{1} \right) \varepsilon_{L}^{\nu} \left( p_{2} \right)
\\
\nonumber
&=& \left[ \frac{1}{m_{W}^{2}} p_{1}^{\mu} p_{2}^{\nu} +
\frac{1}{m_{W}} p_{1}^{\mu} \Delta^{\nu} \left( p_{2} \right)
+ \frac{1}{m_{W}} \Delta^{\mu} \left( p_{1} \right)
p_{2}^{\nu} + \Delta^{\mu} \left( p_{1} \right) \Delta^{\nu}
\left( p_{2} \right) \right] {\cal M}_{\mu \nu}
\end{eqnarray}
The quantities $\Delta \left( p \right)$ occurring in the
second and the third term of (50) can be re-expressed by means of
(38); after a simple manipulation we thus obtain
\begin{equation}
{\cal M}_{LL} =
\left[ - \frac{1}{m_{W}^{2}} p_{1}^{\mu} p_{2}^{\nu} +
\frac{1}{m_{W}} p_{1}^{\mu} \varepsilon^{\nu}_{L} \left( p_{2} \right)
+ \frac{1}{m_{W}} \varepsilon_{L}^{\mu} \left( p_{1} \right)
p_{2}^{\nu} + \Delta^{\mu} \left( p_{1} \right) \Delta^{\nu}
\left( p_{2} \right) \right] {\cal M}_{\mu \nu}
\end{equation}
Employing now the identities (48) and (49), the expression (51)
becomes
\begin{eqnarray}
\nonumber
{\cal M}_{LL} &=& i \frac{p_{1}^{\mu}}{m_{W}} {\cal M}_{\mu 4} +
i \frac{p_{2}^{\nu}}{m_{W}} {\cal M}_{4 \nu} - {\cal M}_{44}
\\
&~& - i \varepsilon^{\mu}_{L} \left( p_{1} \right) {\cal M}_{\mu 4} -
i \varepsilon^{\nu}_{L} \left( p_{2} \right) {\cal M}_{4 \nu} +
\Delta^{\mu} \left( p_{1} \right) \Delta^{\nu} \left( p_{2}
\right) {\cal M}_{\mu \nu}
\\
\nonumber
&=& - {\cal M}_{44} - i \Delta^{\mu} \left( p_{1} \right)
{\cal M}_{\mu 4} - i \Delta^{\nu} \left( p_{2} \right) {\cal
M}_{4 \nu} + \Delta^{\mu} \left( p_{1} \right) \Delta^{\nu}
\left( p_{2} \right) {\cal M}_{\mu \nu}
\end{eqnarray}
For $E \gg m_{W}$, where $E$ is the $W$ boson energy, the
quantities $\Delta \left( p \right)$ are of the order ${\cal
O} \left( m_{W} / E \right)$ and the truncated matrix elements
${\cal M}_{MN}$ in (52) are at most ${\cal O} \left( 1 \right)$.
One may thus finally write, in the high-energy limit
\begin{equation}
{\cal M}_{LL} = - {\cal M}_{44} \left[ 1 + {\cal O} \left(
m_{W}/E \right) \right]
\end{equation}
which is just the statement of the ET for the considered
particular case, including the phase factor $(-1)$ (cf.(27) with
$n_{1} = 0$, $n_{2} = 2$).

Let us add that the result (52) can be generalized in a
straightforward way for processes involving an arbitrary number
of external longitudinal vector bosons. A corresponding relation
(which is implicit in the treatment \cite{10}) has been formulated
explicitly first by H.Veltman (see ref.\cite{Willenbrock}). Recently, it has
been discussed by Grosse-Knetter and Kuss \cite{12} who refer to an
identity of the type (52) as the {\it generalized equivalence theorem.}\/
Following \cite{12}, it should be emphasized that eq.(52) is an {\it
exact\/} relation and the usual ET form of the type (53) is
obtained only when the truncated matrix elements in (52) exhibit
a ``soft'' high-energy behaviour (which is obvious in $R$-gauge
formulation of the SM assuming that Higgs boson $H$ is not much
heavier than the $W$.)

\section {Conclusion}

Concluding these notes, let us return briefly
to the problem of high-energy behaviour of Feynman
graphs within SM as outlined in the Introduction. It is
now clear that with ET at hand, the tree unitarity \cite{3},
\cite{7}
(i.e. the "asymptotic softness" \cite{6} of tree-level amplitudes)
can be proved in two steps: First, one may invoke
gauge-independence of $S$-matrix elements, which is a
consequence of Ward identities of the theory (see e.g. \cite{19}
for a proof within the class of $R_{\xi}$-gauges and \cite{28}
for a discussion of the $U$-gauge limit). By passing from
$U$-gauge to an $R$-gauge, one gets rid of a potential
source of "bad" high-energy behaviour, residing in the
$U$-gauge vector boson propagators (which contain pieces
proportional to $m_V^{-2}$).

Second, having passed to $R$-gauge, one may employ ET
(which also originates in a particular Ward identity) and
this in turn eliminates another possible source of troubles,
namely the vectors of longitudinal polarization (which
contain pieces proportional to $m_V^{-1}$; the unphysical
scalar matrix elements are obviously harmless in the
high-energy limit. In this  way, ET provides remarkable insight
into the nature of the subtle divergence cancellations
characteristic of the SM (as well as of the other non-abelian
gauge theories involving Higgs-Goldstone scalars). It is
clear that the formal ET machinery based on the identities
(43) is substantially more efficient in proving the
tree-level unitarity than straightforward $U$-gauge
calculations, which become rather involved even for relatively
simple processes.

In fact, ET is not only of fundamental importance for proving
some general statements within the electroweak theory, but it
is also often used for simplifying practical calculations of
Feynman diagrams with external massive vector bosons. The
aim of the present lecture notes is to provide some
background for a possible further study of ET and related
topics. As we already mentioned in the Introduction there are
other interesting and important aspects of the
subject, both on the technical side and in the area of
physical applications. However, a corresponding discussion
would go beyond the scope of this introductory treatment.
The current literature concerning ET is rather rich and the
interested reader may find a (presumably incomplete) list e.g.
under ref. \cite{12} -- \cite{Willenbrock},
in addition to the basic references used in
preparing the present text.

\vspace{0.5cm}

\noindent
{\bf Acknowledgement:} I would like to thank Hong-Jian He for
useful comments and correspondence. The work has been supported
in part by the research grants GAUK--166/95 and
GACR--202/95/1460.

\newpage

\noindent

\appendix

{\large \bf Appendix}
\vskip0,5cm
\noindent
{\Large \bf Basics of Standard Model}
\addcontentsline{toc}{section}{\protect\numberline{Basics of
Standard Model}}
\vskip0,5cm
\noindent
Glashow-Weinberg-Salam (GWS) standard model (SM) of electroweak
interactions provides a unification of the parity-violating (V-A)
weak force (responsible e.g. for muon decay and mediated by
charged massive vector bosons $W^{\pm}$) and the parity-conserving
electromagnetic interaction due to the exchange of massless photon
$\gamma$. It is a non-abelian gauge theory where particle masses are
generated via Higgs mechanism.

{\it The gauge group} $SU(2) \times U(1)$ corresponds to four vector
fields $A_{\mu}^a,\; a = 1,2,3$ and $B_{\mu}$. Note that {\it four}
vector bosons are needed since we know that an extra vector boson
(apart from $W^{\pm}$ and $\gamma$) must be introduced in order to
accomplish a technically successful electroweak unification without
any exotic fermions (such as a heavy electron etc.)

{\it The Higgs mechanism} is realized by means of an $SU(2)$ doublet of
complex scalar fields
$$\Phi = \pmatrix{\varphi^+\cr
\varphi^0\cr}\eqno(A.1)$$
(the lower component of (1) is taken to be electrically neutral). It
means that four real scalars are involved; of course, it is so
because one needs, in accordance with general properties of the Higgs
mechanism, three Goldstone bosons in order to get eventually three massive
vector bosons. Note that it would not suffice to take a real
triplet of scalars since such an option would leave us with 2
massless neutral vector bosons - only $W^{\pm}$ would get
a mass in this way. A minimal scalar multiplet is thus
a complex doublet (i.e. real quartet).

The last but not least, there are {\it fermions} (3 generations
of leptons and quarks). In order to describe correctly the parity-violating
weak interactions and the parity-conserving electromagnetism, the
left-handed and right-handed components of fermion fields must
transform differently (doublets for $L$, singlets for $R$).

The $U(1)$ transformation properties of all matter fields (i.e.
scalars and fermions) are defined by the corresponding weak
hypercharge which is given by
$$Q = T_3 + Y \eqno(A.2)$$
($T_3$ is "weak isospin" and $Q$ is electric charge in units of $e$).
As for fermions, we will first restrict ourselves to leptons (electron
type).  Denote simply
$$L = \pmatrix{\nu_L\cr
e_L\cr} \quad, \quad R = e_R \eqno (A.3)$$
Then according to (2) one obviously has
$$Y_L = -{1 \over 2} \quad, \quad Y_R = -1 \eqno(A.4)$$
\vskip0,5cm
\noindent
{\bf The gauge invariant Lagrangian}

The  gauge invariant Lagrangian of the GWS standard model may be
written as consisting essentially of four pieces, namely
$${\cal{L}}_{GWS} = {\cal{L}}_{gauge} + {\cal{L}}_{fermion} +
{\cal{L}}_{Higgs} + {\cal{L}}_{Yukawa} \eqno(A.5)$$
\vskip0,5cm
\noindent
{\bf Interactions of vector bosons with leptons}

Let us begin with ${\cal{L}}_{fermion}$ (for the moment, the
fermions are just $\nu_e$ and $e$). It may be written as
$${\cal{L}}_{fermion} = i \bar{L} \gamma^{\mu} (\partial_{\mu} -
igA_{\mu}^a {\tau^a \over 2} + {i \over 2} g' B_{\mu})L\nonumber\\
+ i \bar{R} \gamma^{\mu} (\partial_{\mu} + ig'B_{\mu})R
\eqno(A.6)$$
where $\tau^a,\; a = 1,2,3$ are Pauli matrices and the $g, \;g'$
are two independent coupling constants. Note that gauge invariance of
the expression (6) is due to the covariant derivatives (by which
we have replaced the ordinary derivatives in the lepton
kinetic terms). In writing (6) we have taken into account the values of weak
hypercharges in (4). Working out the last expression and introducing
the notation
$${W_{\mu}^{\pm}} = {{A_{\mu}^1 \mp iA_{\mu}^2} \over {\sqrt 2}}
\eqno(A.7)$$
one first reproduces the standard charged current weak interaction
$${\cal{L}}_{CC} = {g \over \sqrt 2} \bar \nu_L \gamma^{\mu} q_L
W_{\mu}^+ + {\rm h.c.} \eqno(A.8)$$
Further, in the diagonal part of (6) (i.e. in the terms involving
$\tau^3$ and the unit matrix) neither $A_{\mu}^3$ nor $B_{\mu}$ can be
interpreted as the electromagnetic potential. However, one may
introduce new fields $A_{\mu}$ and $Z_{\mu}$ by means of an
orthogonal transformation
$$A_{\mu}^3 = \cos \vartheta_W Z_{\mu} + \sin \vartheta_W A_{\mu} \qquad
B_{\mu} = -\sin \vartheta_W Z_{\mu} + \cos \vartheta_W A_{\mu}
\eqno(A.9)$$
(let us stress that such  a transformation must be orthogonal so as
not to spoil diagonality of the kinetic energy terms coming from
${\cal{L}}_{gauge}$). Using (9) in eq. (6) one is able to
reproduce a correct electromagnetic interaction (i.e. such that
does not involve the $\gamma_5$ and the neutrino field) by choosing the
mixing angle $\vartheta_W$ (the Weinberg angle) to satisfy
$${\cos \vartheta_W} = {{g} \over {\sqrt {g^2 + g'^2}}} \quad, \quad
{\sin \vartheta_W} = {{g'} \over {\sqrt {g^2 + g'^2}}} \eqno(A.10)$$
The electromagnetic coupling constant $e$ then comes out to be
$$e = {{gg'} \over {\sqrt {g^2 + g'^2}}} = g \sin \vartheta_W \eqno(A.11)$$
One thus gets an important constraint on the relative strength of charged-current
weak interaction ($g$) with respect to the electromagnetic coupling.
From $e = g \sin \vartheta_W$ one has
$$e < g  \eqno(A.12)$$
(note that $e = g$ is excluded as it is not compatible with (10) for
$g \not= 0$). The relation (11) or (12) resp. is sometimes called
a "unification condition" in the literature. An important consequence
of eq. (11) is a lower bound for the $W$ mass. Indeed, taking
into account the well known relation
$${G_F \over \sqrt 2} = {g^2 \over 8m_W^2} \eqno(A.13)$$
(which expresses compatibility of the Fermi-type theory of weak interactions
with a model involving the $W$ boson) then using (11) and the
definition of the fine structure constant $\alpha = e^2/4 \pi$,
one gets the result
$$m_W = \biggr({\pi \alpha \over G_F \sqrt 2} \biggr)^{1 \over 2}
{1 \over \sin \vartheta_W} \eqno(A.14)$$
Taking in eq. (14) $G_F \doteq 1.166 \times 10^{-5} GeV^{-2}$ and
$\alpha \doteq 1/137$, one gets immediately a lower bound
$$m_W > 37\; \rm GeV \eqno(A.15)$$
It is remarkable that one is able to derive the result (14) leading
to the estimate (15) without even mentioning the Higgs mechanism;
thus Glashow in his 1961 paper could in fact predict such a
lower bound for $m_W$ (but he failed to do so).

For an interaction of the $Z$ boson with leptons (i.e. for a weak
neutral current interaction) one then gets the result (a GWS
{\it prediction})
$${\cal{L}}_{NC} = {g \over \cos \vartheta_W} \biggr[ {1 \over 2}
\bar \nu_L \gamma^{\mu} \nu_L + \biggr( -{1 \over 2} + \sin^2
\vartheta_W \biggr) \bar e_L \gamma^{\mu} e_L + \sin^2
\vartheta_W \bar e_R \gamma^{\mu} e_R \biggr] Z_{\mu}
\eqno(A.16)$$
Note that the coefficients of the individual terms in the square
brackets in (16) satisfy a simple rule
$$\varepsilon_f = T_{3f} - Q_f \sin^2 \vartheta_W \eqno(A.17)$$
(where $f$ stands for $\nu_L,\; e_L$ or $e_R$).

Of course, now one should also add the other lepton types $\mu$
and $\tau$ (which is trivial if a possible lepton mixing is
ignored) and, moreover, one has to include 3 generations of quarks. The
incorporation of quarks will be described briefly somewhat later.
Here let us emphasize that the sector of lepton-vector boson
interactions is nowadays the best tested part of SM.
\vskip0,5cm
{\bf Vector boson self-interactions}

Another part of the SM Lagrangian which is generally considered
quite trustworthy at present (although its precise experimental
tests still lie ahead of us) is the sector of vector boson
self-interactions, i.e. the term ${\cal{L}}_{gauge}$ in (5). Let
us now summarize some familiar facts concerning the construction of this
sector. The ${\cal{L}}_{gauge}$ may be written as
$${\cal{L}}_{gauge} = - {1 \over 4} F_{\mu \nu}^a F^{a \mu \nu} -
{1 \over 4} B_{\mu \nu} B^{\mu \nu} \eqno(A.18)$$
where
$$F_{\mu \nu}^a = \partial_{\mu} A_{\nu}^a - \partial_{\nu} A_{\mu}^a +
g \varepsilon^{abc} A_{\mu}^b A_{\nu}^c \eqno(A.19)$$
$$B_{\mu \nu} = \partial_{\mu} B_{\nu} - \partial_{\nu} B_{\mu}
\eqno(A.20)$$
Let us remark that gauge invariance of the non-abelian part of (18)
is most transparent if one employs the definition of the $F_{\mu \nu}^a$
in terms of a commutator of $SU(2)$ covariant derivatives, namely
$$- ig F_{\mu \nu} = [ D_{\mu},\; D_{\nu}] \eqno(A.21)$$
where
$$F_{\mu \nu} = F_{\mu \nu}^a T^a \eqno(A.22)$$
and
$$D_{\mu} = \partial_{\mu} - ig A_{\mu} \eqno(A.23)$$
with $A_{\mu} = A_{\mu}^a T^a$; the $T^a$ are $SU(2)$ generators
(e.g. $T^a = {1 \over 2} \tau^a, \; a = 1,2,3$). The first term in
eq. (18) may be then recast as
$$- {1 \over 4} F_{\mu \nu}^a F^{a \mu \nu} = - {1 \over 2} {\rm Tr} (F_{\mu \nu}
F^{\mu \nu}) \eqno(A.24)$$
(taking into account $Tr (T^a T^b) = {1 \over 2} \delta^{ab}$)
and this makes the gauge invariance obvious, since the $F_{\mu \nu}$ is
transformed covariantly owing to (21).

While the abelian part of (18) is of course just a kinetic term for
the $U(1)$ gauge field $B_{\mu}$, the non-abelian part produces, beside
kinetic terms for the $A_{\mu}^a$, also some specific self-interactions
(trilinear and quadrilinear in gauge potentials). When the expression
(18) is recast in terms of physical vector boson fields $W_{\mu}
^{\pm}, \; Z_{\mu}$ and $A_{\mu}$ (see (7) and (9)) one gets
$${\cal{L}}_{gauge} = - {1 \over 2} W_{\mu \nu}^- W^{+ \mu \nu} -
{1 \over 4} Z_{\mu \nu} Z^{\mu \nu} - {1 \over 4} A_{\mu \nu}
A^{\mu \nu}
+ {\cal{L}}_{VVV} + {\cal{L}}_{VVVV} \eqno(A.25)$$
where the notation in the kinetic terms should be self-explanatory and
the interactions have the following form (the $V$ is a generic
symbol for any vector boson, i.e. $W^{\pm}, \; Z$ or $\gamma$)
$${\cal{L}}_{VVV} = - ig (W_{\mu}^0 W_{\nu}^- \overleftrightarrow{\partial^{\mu}}
W^{+ \nu} + W_{\mu}^- W_{\nu}^+ \overleftrightarrow{\partial^{\mu}} W^{0 \nu} +
W_{\mu}^+ W_{\nu}^0 \overleftrightarrow{\partial^{\mu}} W^{- \nu})
\eqno(A.26)$$
$$
{\cal{L}}_{VVVV} = - g^2 \biggr[ {1 \over 2} (W^- . W^+)^2 -
{1 \over 2} (W^-)^2 (W^+)^2 +(W^- . W^+) (W^0)^2
$$
$$
\phantom{{\cal{L}}_{V}\,\;\;  \biggr[ {1 \over 2} ( W^+)^2 }
- (W^- . W^0) (W^+ . W^0)\biggr] \eqno(A.27)
$$
where $W_{\mu}^0 = A_{\mu}^3 = \cos \vartheta_W Z_{\mu} + \sin
\vartheta_W A_{\mu}$, and an obvious shorthand notation for Lorentz
scalar products has been used in (27). It is interesting to observe
that the expressions (26) and (27) comprise just some particular
types of the vector boson interactions, namely the triple vector
boson couplings $WW \gamma$ and $WWZ$ and the quartic couplings $WWWW,\;
WWZZ,\; WWZ \gamma$ and $WW \gamma \gamma$. For instance, there
are no direct triple interactions like $ZZZ$ etc. (technically this
is due to antisymmetry of the $SU(2)$ structure constants
$f^{abc} = \varepsilon^{abc}$) and there are no direct quartic couplings
like $Z \gamma \gamma \gamma$ etc. (these may be induced in higher
orders).

\vskip0,5cm
\noindent
{\bf Higgs sector}

The least understood part of the SM is its "Higgs sector" or, in
other words, the sector responsible for the electroweak symmetry
breaking and for generating particle masses. Let us start with the
term ${\cal{L}}_{Higgs}$ in (5). This is obtained by replacing
ordinary derivatives in scalar kinetic energy by covariant
derivatives and it also includes a "potential" of the Goldstone
type, i.e.
$${\cal{L}}_{Higgs} = (D^{\mu} \Phi)^{\dagger} (D_{\mu} \Phi) - V(\Phi)
\eqno(A.28)$$
where
$$D_{\mu} = \partial_{\mu} - ig A_{\mu}^a {\tau^a \over 2} -
{i \over 2} g' B_{\mu} \eqno(A.29)$$
(notice that in (29) we have used $Y_{\Phi} = {1 \over 2}$). The
"potential" $V(\Phi)$ has a familiar form
$$V(\Phi) = - \mu^2 \Phi^{\dagger} \Phi + \lambda (\Phi^{\dagger} \Phi)^2
\eqno(A.30)$$
The $V$ is minimized for constant $\Phi_0$ such that
$$\Phi_0^{\dagger} \Phi_0 = {v^2 \over 2} \eqno(A.31)$$
where
$$v = \mu / \sqrt{\lambda} \eqno(A.32)$$
The $\Phi$ may be written as (a simple exercise in matrix multiplication)
$$\Phi = e^{i \xi^a \tau^a}
 {0 \choose {\frac{1}{\sqrt{2}}(v+H)}} \eqno(A.33)$$
where $H$ is (massive) Higgs boson field $(m_H^2 = 2 \lambda v^2)$
and the $\xi^a, \; a= 1,2,3$ represent the would-be Goldstone bosons.
These can be gauged away, setting thus effectively $\xi^a = 0$; this is
equivalent to the corresponding $SU(2)$ gauge transformation
$$\Phi \to \Phi^{(U)} =
 {0 \choose {\frac{1}{\sqrt{2}}(v+H)}} \eqno(A.34)$$
fixing the so-called unitary gauge (or simply $U$-gauge).

From ${\cal{L}}_{Higgs}$ one then gets first of all mass terms for the
$W^{\pm}$ and $Z$ bosons (by combining the constant shift $v$ of the
Higgs field with the gauge fields from covariant derivatives). The
resulting values of the $W$ and $Z$ masses are
$$m_W = \frac {1}{2} gv \; , \qquad
m_Z = \frac {1}{2}  (g^2 +g'^2)^{1/2} v \eqno(A.35)$$
i.e. one has
$${m_W}/{m_Z} = \cos \vartheta_W \eqno(A.36)$$
Note that the existence of such a relation is closely related to the
fact that $\Phi$ is a $SU(2)$ {\it doublet}. Let
us also remark that from $m_W = \frac{1}{2} gv$ and from
$G_F / \sqrt 2 = g^2/ 8 m^2_W$ one may get easily
$$v = (G_F \sqrt 2)^{-1/2} \doteq 246\; \rm GeV \eqno(A.37)$$

Further, using (34) in ${\cal{L}}_{Higgs}$ one gets interactions
of the type  $WWH,\; ZZH$, $WWHH$ and $ZZHH$ as well as Higgs boson
self-interactions (cubic and quartic). The relevant coupling
constants are, for example
$$g_{WWH} = gm_W \eqno(A.38)$$

$$g_{WWHH} = \frac{1}{4}g^2 \eqno(A.39)$$

$$g_{HHHH} = - \frac{1}{4} \lambda, \qquad
\lambda = \frac {G_F m_H^2} {\sqrt 2} \eqno(A.40)$$

The  last but not least, there is a Yukawa-type interaction of the
Higgs doublet $\Phi$ with leptons. It may be written in an obviously
$SU(2)$ invariant form as
$${\cal{L}}_{Yukawa} = - h_e \bar L \Phi R + \mbox{h.c.} \eqno(A.41)$$
(it is not difficult to verify that the last
expression is also $U(1)$ invariant). The
$h_e$ is a coupling constant which may be easily related to other
relevant physical parameters. Indeed, using (34) in (41) one gets
immediately a mass term for the electron, with
$$m_e = h_e \frac {v} {\sqrt 2} \eqno(A.42)$$
Similarly one obtains a scalar Yukawa interaction of the Higgs boson
with electron
$${\cal{L}}_{eeH} = g_{eeH} \bar e e H \eqno(A.43)$$
with
$$g_{eeH} = - \frac{g}{2} \frac {m_e}{m_W} \eqno (A.44)$$
(in arriving at (44) one has to use (42) and take into account that
$v = 2 m_W / g$ - see (35)). Eq. (44) thus expresses the well-known
dependence of the $H$ couplings to fermions on fermion masses,
characteristic for the minimal SM Higgs sector.

\vskip0,5cm
\noindent
{\bf{Full fermionic sector of the standard model}}

The present-day SM incorporates 3 generations of leptons and
quarks. For simplicity, in what follows we will neglect a
possible mixing in the lepton sector as well as neutrino masses.
The basic ``building blocks'' of the lepton sector are
then the following:

\noindent
$SU(2)$ doublets
\begin{eqnarray}
\nonumber
\left( \matrix{ & \nu_{e} \cr
& e \cr} \right)_{L}, \quad
\left( \matrix{ & \nu_{\mu} \cr
& \mu \cr} \right)_{L}, \quad
\left( \matrix{ & \nu_{\tau} \cr
& \tau \cr} \right)_{L} \quad : \quad
\mbox{weak hypercharge $Y^{(l)}_{L} = - \frac{1}{2}$ }
\end{eqnarray}

\noindent
$SU(2)$ singlets
\begin{eqnarray}
\nonumber
e_{R}, \quad \mu_{R}, \quad \tau_{R} \quad : \quad
\mbox{weak hypercharge $Y^{(l)}_{R} = -1$}
\end{eqnarray}

The quark sector is built as follows:

\noindent
$SU(2)$ doublets
\begin{eqnarray}
\nonumber
\left( \matrix{ & u_{0} \cr
& d_{0} \cr} \right)_{L}, \quad
\left( \matrix{ & c_{0} \cr
& s_{0} \cr} \right)_{L}, \quad
\left( \matrix{ & t_{0} \cr
& b_{0} \cr} \right)_{L} \quad : \quad
\mbox{weak hypercharge $Y^{(u)}_{L} =  \frac{1}{6}$ }
\end{eqnarray}

\noindent
$SU(2)$ singlets

$u_{0 \, R}, \quad c_{0 \, R}, \quad t_{0 \, R} \quad : \quad
\mbox{weak hypercharge $Y^{(u)}_{R} = \frac{2}{3}$}$

$d_{0 \, R}, \quad s_{0 \, R}, \quad b_{0 \, R} \quad : \quad
\mbox{weak hypercharge $Y^{(d)}_{R} = - \frac{1}{3}$}$

\vspace{2mm}
\noindent
The quark fields labelled by a subscript zero are not, in
general, identical with mass eigenstates. Physical quark masses
arise from general Yukawa interactions similarly to the lepton
case; however, since all quarks are a~priori assumed to be
massive, the Yukawa couplings produce a general mass matrix that
has to be diagonalized in order to define the mass eigenstates.
Such a diagonalization provides in turn a natural description of
the mixing in quark sector.

When dealing with the quark sector it is important to notice
first that in order to give masses both to $u$-type quarks (with
charge $2/3$) and to $d$-type ones (with charge $-1/3$) via Yukawa
interactions one has to employ, beside the Higgs doublet $\Phi$,
also a conjugate object $\widetilde{\Phi}$ defined as
$$\widetilde{\Phi} = i \tau_{2} \Phi^{\ast} \eqno(A.45)$$
(prove that (45) is indeed a doublet with respect to
$SU(2)$). Introducing a general Yukawa-type interaction for
quarks, then through the $\Phi$ one gets a general mass matrix
(not necessarily hermitean) for $d$-type quarks
$${\cal{L}}_{mass} = - ( \overline{d_{0 \, L}}, \;
\overline{s_{0 \, L}}, \; \overline{b_{0 \, L}} ) M \left(
\matrix{ & d_{0 \, R} \cr & s_{0 \, R} \cr & b_{0 \, R} \cr}
\right) + \mbox{h.c.}\eqno(A.46)$$
and analogously (through the $\widetilde{\Phi}$) another mass
matrix $\widetilde{M}$ for $u$-type quarks.

Now, every non-singular complex square matrix $M$ or
$\widetilde{M}$ may be diagonalized by means of a
{\it{biunitary}} transformation, i.e.
$$M = {\cal{U}}^{\dagger} {\cal{M}} {\cal{V}} \eqno(A.47)$$
$${\widetilde{M}} = {\widetilde{\cal{U}}}^{\dagger} {\widetilde{\cal{M}}}
{\widetilde{\cal{V}}}$$
where ${\cal{U}}$, ${\cal{V}}$ and $\widetilde{\cal{U}}$,
$\widetilde{\cal{V}}$ are unitary matrices and $\cal{M}$,
$\widetilde{\cal{M}}$ are diagonal positive definite. Thus, the
left-handed components of the ``mass eigenstates'' may be
written as
$$\left( \matrix{ & u_{L} \cr & c_{L} \cr & t_{L} \cr} \right) =
{\widetilde{\cal{U}}} \left( \matrix{& u_{0 \, L} \cr & c_{0 \,
L} \cr & t_{0 \, L} \cr} \right), \quad
\left( \matrix{ & d_{L} \cr & s_{L} \cr & b_{L} \cr} \right) =
{\cal{U}} \left( \matrix{& d_{0 \, L} \cr & s_{0 \,
L} \cr & b_{0 \, L} \cr} \right) \eqno(A.48)$$
(and similarly for the right-handed components).

The charged-current quark weak interactions involve only
left-handed quarks, namely
$${\cal{L}}^{\sixrm{(quark)}}_{\sixrm{CC}} = \frac{g}{\sqrt{2}} (
\overline{u_{0 \, L}}, \; \overline{c_{0 \, L}}, \;
\overline{t_{0 \, L}} ) \gamma^{\mu} \left( \matrix{& d_{0 \, L}
\cr & s_{0 \, L} \cr & b_{0 \, L} \cr} \right) W^{+ \mu} +
\mbox{h.c.} \eqno(A.49)$$
The last expression may be now recast in terms of the ``mass
eigenstates''; using the transformation (48) in
(49) one gets
$${\cal{L}}^{\sixrm{(quark)}}_{\sixrm{CC}} = \frac{g}{\sqrt{2}} (
\overline{u_{L}}, \; \overline{c_{L}}, \;
\overline{t_{L}} ) \gamma^{\mu}
{\widetilde{\cal{U}}}{\cal{U}}^{\dagger} \left( \matrix{& d_{L}
\cr & s_{L} \cr & b_{L} \cr} \right) W^{+ \mu} +
\mbox{h.c.} \eqno(A.50)$$
The matrix ${\widetilde{\cal{U}}}{\cal{U}}^{\dagger}$ appearing
in (50) may of course be identified with the famous
Cabibbo---Kobayashi---Maskawa matrix $V_{\mbox{\sixrm{CKM}}}$.
(Note that the result (50) makes it clear that it would
not be physically meaningful to consider a mixing of the
$u$-type and $d$-type quarks separately; thus the CKM mixing is
conventionally assumed to occur among $d$-type quarks.) Recall
that the unitary $3 \times 3$ matrix in (50) contains in
fact just four physically relevant real parameters, which may be
interpreted as three Cabibbo-like angles $\vartheta_{i},
i=1,2,3$ and one phase $\delta$. As it is well known, a possible
non-zero value of the phase $\delta$ is supposed to be a source
of the CP violation within the standard model. As an instructive
exercise for the interested reader, we leave it to prove that in
the case of $n$ generations a generalized CKM matrix would
involve $(n-1)^{2}$ physically relevant real parameters
($\frac{1}{2} n (n-1)$ angles plus $\frac{1}{2} (n-1) (n-2)$
phases).

In particular, if $n=2$, one obtains the familiar
Glashow---Iliopoulos---Maiani mixing matrix
$$V_{\mbox{\sixrm{GIM}}} = \left( \matrix{ & \cos \vartheta_{C} &
\sin \vartheta_{C} \cr
& - \sin \vartheta_{C} & \cos \vartheta_{C} \cr} \right) \eqno(A.51)$$
where $\vartheta_{C}$ is the Cabibbo angle.

The good news concerning the above-described picture is that quark
weak {\it{neutral currents}} remain flavour-diagonal (owing to
${\cal{U}} {{\cal{U}}^{\dagger}} = 1$ etc.) --- i.e. the
(generalized) GIM mechanism works (needless to say, the
electromagnetic quark current is flavour-diagonal as well). It
is good news indeed since otherwise one would run into
phenomenological disaster caused by the notorious ``flavour
changing neutral currents'' (FCNC). At present there are
numerous experimental data showing that flavour-changing weak
decays conserving the hadronic charge are strongly suppressed in
comparison with the corresponding charge-changing decays, so
that FCNC interactions must be absent or strongly suppressed.
Within the standard model, FCNC are absent at the tree level as
we have already remarked; they can only be induced at one-loop
(or higher) level and one may thus understand (even
quantitatively, to some extent) the corresponding suppression
factors for some typical processes.

To close this section let us add that along with the
above-mentioned diagonalization of the quark mass matrices, the
Yukawa interactions are dia\-go\-na\-li\-zed as well and one gets a
pattern of Higgs-quark couplings which is completely analogous
to the lepton case. Within SM one thus has, for any
fermion--Higgs interaction (cf. (44))
$${\cal{L}}_{ffH} = - \frac{g}{2} \frac{m_{f}}{m_{W}} {\overline{f}}
f H \eqno(A.52)$$

\end{document}